\documentclass[useAMS,usenatbib]{mn2e}

\usepackage{times}
\usepackage{natbib}
\usepackage{color}
\usepackage{todonotes}
\usepackage{siunitx}
\usepackage{url}
\usepackage{hyperref}
\usepackage{amssymb}

\newcommand{\apjs}{ApJS}

\newcommand{\apjl}{ApJL}

\newcommand{\aap}{A \& A}

\DeclareGraphicsExtensions{.pdf,.png}

\begin{document}
 
\title[Radio U-Net]{Radio U-Net: a convolutional neural network to detect diffuse radio sources in galaxy clusters and beyond}
\author[C. Stuardi et al.]{C. Stuardi$^1$\thanks{E-mail: ccstuardi@gmail.com}, C. Gheller$^1$, F. Vazza$^{2,1,3}$, A. Botteon$^1$\\
$^{1}$Istituto di Radio Astronomia, INAF, Via P. Gobetti 101, 40129 Bologna, Italy\\
$^{2}$ Dipartimento di Fisica e Astronomia, Universit\'{a} di Bologna, Via P. Gobetti 92/3, 40129 Bologna, Italy\\
$^{3}$ Hamburger Sternwarte, Gojenbergsweg 112, 21029 Hamburg, Germany
}

\date{Accepted 2024 August 20. Received in original form 2024 June 24}
\maketitle
\begin{abstract}
The forthcoming generation of radio telescope arrays promises significant advancements in sensitivity and resolution, enabling the identification and characterization of many new faint and diffuse radio sources. Conventional manual cataloging methodologies are anticipated to be insufficient to exploit the capabilities of new radio surveys. Radio interferometric images of diffuse sources present a challenge for image segmentation tasks due to noise, artifacts, and embedded radio sources. In response to these challenges, we introduce Radio U-Net, a fully convolutional neural network based on the U-Net architecture. Radio U-Net is designed to detect faint and extended sources in radio surveys, such as radio halos, relics, and cosmic web filaments. Radio U-Net was trained on synthetic radio observations built upon cosmological simulations and then tested on a sample of galaxy clusters, where the detection of cluster diffuse radio sources relied on customized data reduction and visual inspection of LOFAR Two Metre Sky Survey (LoTSS) data. The $83\%$ of clusters exhibiting diffuse radio emission were accurately identified, and the segmentation successfully recovered the morphology of the sources even in low-quality images. In a test sample comprising 246 galaxy clusters, we achieved a $73\%$ accuracy rate in distinguishing between clusters with and without diffuse radio emission. Our results establish the applicability of Radio U-Net to extensive radio survey datasets, probing its efficiency on cutting-edge high-performance computing systems. This approach represents an advancement in optimizing the exploitation of forthcoming large radio surveys for scientific exploration.
\end{abstract}

\label{firstpage} 

\section{Introduction}
\label{sec:intro}

The backbone of the universe is the cosmic web: an intricate network of filaments interconnected with galaxy clusters. The thermal plasma that fills filaments (the Warm Hot Intergalactic Medium, WHIM) and galaxy clusters (the Intra Cluster Medium, ICM) is permeated by weak magnetic fields (0.01-10 $\mu$G) whose origin and evolution have attracted increasing attention in the scientific community in recent years \citep[see e.g.][]{Vazza21,Carretti23}. Structure formation processes inject shocks and turbulence into the thermal plasma, accelerating particles to ultra-relativistic energies. Cosmic ray electrons, spinning in the magnetic fields, emit radio synchrotron radiation in the form of diffuse radio sources which are observed in galaxy clusters \citep{vanWeeren19}, or in filaments and bridges within them \citep{Govoni19,Botteon20c,Vernstrom21}. Diffuse radio sources are primary probes for investigating cosmic magnetic fields, particle acceleration mechanisms, and the processes driving the formation of large-scale cosmic structures.

Diffuse radio sources in galaxy clusters have historically been categorized into giant radio halos, mini-halos, and radio relics based on their location in the cluster, morphology, size, polarization, and spectral characteristics \citep{Feretti12} However, the boundaries between the different classes are not always clear-cut. For instance, the discovery of Mpc-size sources in clusters hosting mini-halos suggests the presence of an intermediate stage between mini-halos and giant radio halos \citep[e.g.,][]{Kale19, Savini19, Biava21}, leading to the grouping of these two classes into the common category of radio halos. Furthermore, diffuse radio sources often have unique morphologies \citep[e.g.,][]{deGasperin17,Botteon20b}, and projection effects may cause the overlap of different sources along the line-of-sight \citep[e.g.,][]{Rajpurohit20a}. Sometimes diffuse radio sources are connected with radio galaxies or revived fossil plasma sources, also known as phoenices (see \citealt{Mandal20} and \citealt{Raja23} for some recent works) which can furnish them with a reservoir of mildly relativistic electrons.

Diffuse radio emission has recently been detected extending into the outskirts of galaxy clusters in a few systems. In some cases, the radio emission fills the volume of galaxy clusters on large scales \citep{Shweta20, Cuciti22, Rajpurohit21, Botteon22b}. This discovery impacts our understanding of particle acceleration efficiency and diffusion within the turbulent ICM on very broad scales \citep{Beduzzi23, Nishiwaki24}. Radio emission has been detected even between interacting galaxy clusters, in the form of bridges with projected lengths of 1-2 Mpc  \citep{Govoni19, Botteon20c,Bruno23}. The underlying acceleration mechanism remains a topic of debate \citep{Brunetti20, Vernstrom23}. While stacking experiments recently claimed the detection of radio emission from 15 Mpc-long filaments \citep{Vernstrom21}, the direct detection of synchrotron emission from the cosmic web has not yet been achieved, leaving many questions about the origins of cosmic magnetic fields unanswered.

The common observational feature of all diffuse radio sources, both within and outside galaxy clusters, is that they are rare (fewer than 200 so far) and extended sources with low surface brightness ($\leq$ 1~$\mu$Jy/arcsec$^2$ at 1.4 GHz). Current radio interferometers, like the Low Frequency Array\footnote{\url{https://www.astron.nl/telescopes/lofar/}} \citep[LOFAR,][]{2013A&A...556A...2V}, MeerKAT\footnote{\url{https://www.sarao.ac.za/science/meerkat/about-meerkat/}} \citep{Jonas16}, the Murchison Widefield Array\footnote{\url{https://www.mwatelescope.org/}} \citep[MWA,][]{Tingay13}, and the Australian Square Kilometre Array Pathfinder\footnote{\url{https://www.atnf.csiro.au/projects/askap/index.html}} \citep[ASKAP,][]{Hotan21}, are pushing the boundaries of low-surface brightness detection, discovering new faint and diffuse radio emissions that were previously invisible. Diffuse radio sources also have a steep spectrum: $\alpha>1$, with $S_{\nu} \propto \nu^{-\alpha}$, where $S_{\nu}$ is the flux density at frequency $\nu$ and $\alpha$ is the spectral index. Hence, they are brighter at low radio frequencies. Not surprisingly, the largest compilation of diffuse cluster radio sources to date has been created with LOFAR, at frequencies below 200 MHz \citep{Botteon22}. This compilation comprises 83 clusters that host a radio halo and 26 that host one or more radio relics. While other compilations of diffuse radio sources in galaxy clusters are rapidly growing in number and size \citep{vanWeeren21, Knowles22, Duchesne21b, Duchesne24}, the scientific community looks for new methodologies to detect and classify such objects in current and forthcoming large sky surveys. A further increase in the number of detected diffuse radio sources, within clusters and beyond, is expected during the next decade with the advent of the Square Kilometre Array \citep[SKA\footnote{\url{https://www.skatelescope.org/}}, see e.g.][]{Paul23}.

The complete exploitation of this goldmine of data is essential for extending and validating theoretical studies based on the statistical properties of large samples of objects, particularly for sources in galaxy clusters \citep[e.g.,][]{Botteon20a, Stuardi22, Cassano23, Cuciti23, Jones23, Biava24}. It is also crucial for detecting more sources of diffuse radio emission outside galaxy clusters and shedding light on their origin.

Data obtained by radio interferometers are processed with the aperture synthesis technique to obtain images of the sky. This procedure includes the Fourier transform of the signal collected and it is highly computationally expensive. In addition to the thermal noise of the instrument, interferometric images can be affected by imaging artifacts arising from the antenna's configuration (e.g., missing baselines), ionospheric disturbances, or radio frequency interference that may complicate the calibration and imaging processes. The detection of diffuse radio sources with low surface brightness requires minimizing all these effects and includes the meticulous subtraction of all uncorrelated radio sources. Furthermore, the classification of these sources requires multi-frequency data. For example, comparison with optical images is essential to exclude connections to radio-active galaxies, while X-ray data is used to assess the position of the radio source with respect to the thermal plasma.

The scenario will be further complicated by the overwhelming volume of data that upcoming instruments will soon generate. The SKA Observatory is expected to archive more than 700 petabytes of data per year. Even the current generation of radio telescopes, such as LOFAR, MeerKAT, MWA, and ASKAP, are already creating extensive surveys that push the boundaries of available hardware and software capabilities in terms of data processing and storage. Consequently, future software solutions must be equipped to handle these vast datasets. This necessitates two key developments: 1) the software must efficiently exploit the power of High-Performance Computing (HPC) systems, including parallel processing, diverse hardware configurations, and complex memory and storage hierarchies; and 2) the software must be automated, to reduce human intervention, as manual data processing would be impractical.

A possible way to develop such a new data processing methodology is with the use of deep learning (DL), a branch of artificial intelligence developed for image processing and analysis tasks, like semantic segmentation and classification. DL has already been extensively used in astronomy, astrophysics, and cosmology \citep[see e.g.,][]{2017arXiv170705167S,2017arXiv170906257M,2017arXiv170905889N,2017MNRAS.472.3101G,2018arXiv180106381H,2018arXiv180107615H,2018arXiv180303084C,2018MNRAS.tmp..588A,2018MNRAS.476.1151P,2021arXiv210607660S,2021arXiv210601571C,2021arXiv210412980F}. Specifically relevant to our work is the use of DL for the detection and classification of sources in radio images. In recent years, considerable effort has been dedicated to radio galaxies, for which we already have vast and complete databases that can be used to train supervised networks \citep{2017ApJS..230...20A, 2018MNRAS.476..246L, Lao23, Alegre24, 2023A&C....4200682R, Gupta24}, as well as self-supervised models \citep{Mostert21, Slijepcevic24, Riggi24}. When the dataset size is not sufficient for training a DL model, transfer learning can be exploited to transfer knowledge learned from a large dataset to a smaller one. This method is widely used to address astrophysical classification problems \citep{Ackermann18, 2021arXiv210600187K,Tan23}. However, the success of this process heavily relies on the availability of data for fine-tuning the model to the new task and on the similarity between the original and new datasets.

DL has also been exploited for the detection of sources in our previous work \citep[][]{2018MNRAS.480.3749G}, where we have explored the potential of Convolutional Neural Networks (CNN) in identifying the faint signal from diffuse radio sources in noisy radio observations, which are at the limit of sensitivity of instruments like ASKAP or LOFAR  \cite[see also][]{2022MNRAS.509..990G}. In this work, we trained and tested our models on synthetic radio observations based on cosmological simulations, including both random and systematic perturbations related to the telescope and/or the environment. The resulting methodology, named COSMODEEP, allowed us to detect diffuse radio sources and to localize their position within large images, thanks to a tiling-based procedure with an accuracy of around 90\% for synthetic observations (i.e. around 10\% of the detections are missed). However, the COSMODEEP algorithm is unable to perform accurate image segmentation, which prevents it from providing a precise outline of the shape and extent of the detected object. This limitation hinders further characterization of the detected source, such as performing a morphological classification of the object. 

We aim to overcome this limitation by enhancing our methodology with a fully CNN approach, using the U-Net architecture. This approach has been demonstrated to be extremely effective for image segmentation tasks \citep[e.g.,][]{Ronneberger15}. It is interesting to note that our DL model has been developed to perform the segmentation of data from a completely different scientific application, i.e. photoelastic images from experiments of granular materials \citep{Sanvitale2022}. However, it proves to be effective for our astronomical data, highlighting the flexibility and the strength of DL methods.

In this work, we also aim to extend the application of our methodology to real observations of diffuse radio emissions in galaxy clusters, specifically targeting LOFAR data. However, despite the increasing number of cluster diffuse radio sources discovered in recent years, their quantity is still insufficient for properly training a DL network with real data. To address this problem, we used synthetic observations based on cosmological simulations to train the network to recognize the morphology of the sources even below thermal noise and imaging artifacts. We then investigated the application of the network to LOFAR observations and tested the effectiveness of transfer learning by fine-tuning the pre-trained network on actual observations.

The goal of this work is to provide a first use case of our network by testing its ability to detect diffuse radio sources in a previously known sample of galaxy clusters. Having such a result in hand will enable us to use Radio U-Net for blind searches of diffuse radio emission on the entire sky.
 
The paper is organized as follows. The architecture of Radio U-Net is illustrated in Sec.~\ref{sec:unet}. LOFAR and synthetic observations are described in Sec.~\ref{sec:lofar} and Sec.~\ref{sec:sinth}, respectively. The training and validation of the network on synthetic observations are explained in Sec.~\ref{sec:trainandval} while the results obtained from the application of Radio U-Net to LOFAR data are described in Sec.~\ref{sec:results}. The results are discussed in Sec.~\ref{sec:discussion} and final conclusions are drawn in Sec.~\ref{sec:conclusions}.

\begin{figure*}
\begin{center}
\includegraphics[width=0.8\textwidth]{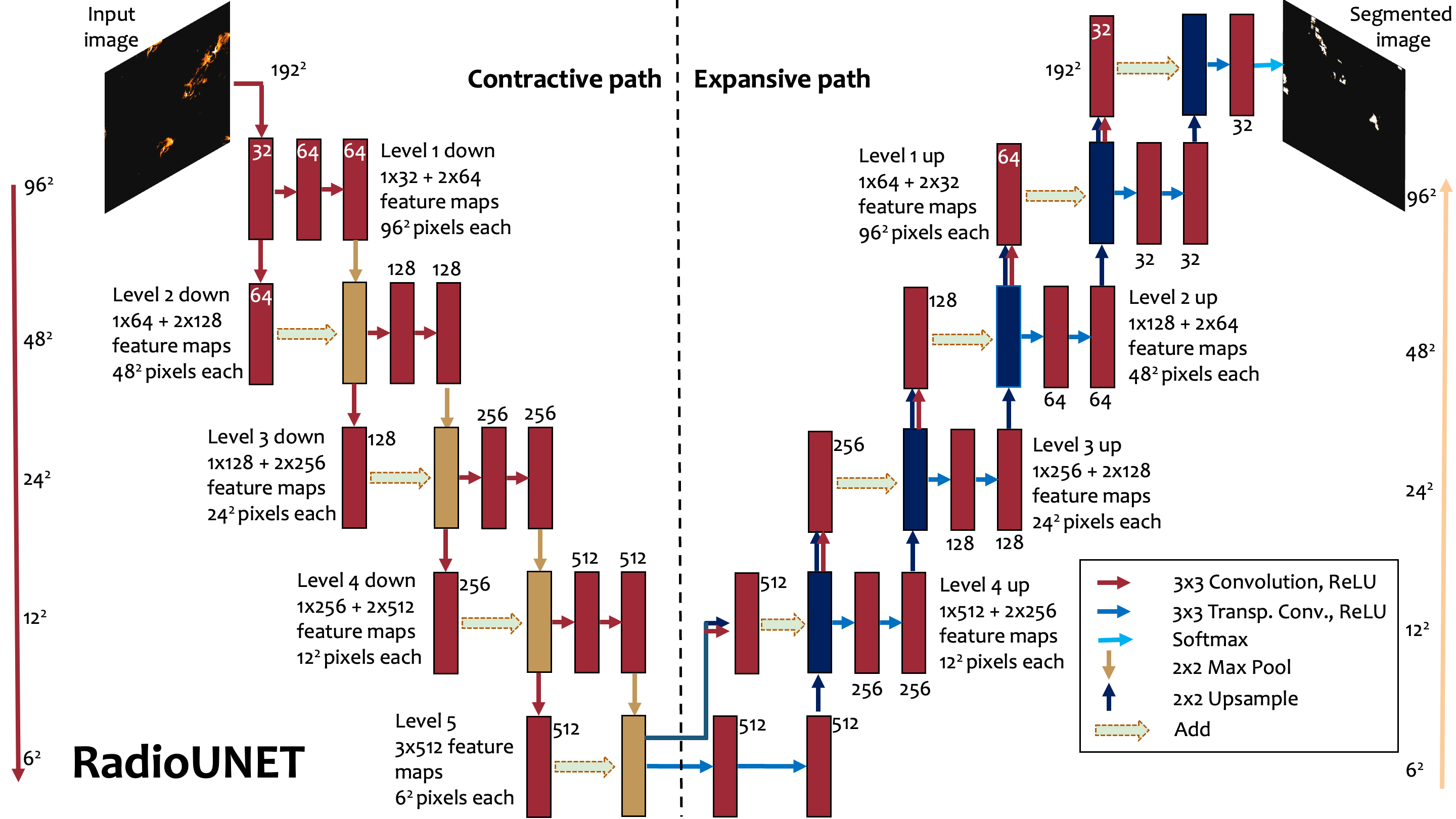}
\caption{Schematic representation of the Radio U-Net architecture.}
\label{fig:base_architecture}
\end{center}
\end{figure*}

\section{The Convolutional Neural Network: Radio U-Net}
\label{sec:unet}

We have adapted the original U-Net architecture \citep{Ronneberger15} to our problem. Fig.~\ref{fig:base_architecture} shows an illustrative representation of the Radio U-Net network. Radio U-Net is an autoencoder consisting of a contractive and an expansive path. The contractive path is composed of a downsampling convolutional network which starts from an input layer, loading the input images, and ends with the deepest convolutional layers. The expansive path starts from this deepest level, followed by an upsampling convolutional network, specular to the contractive one. At each level, the feature maps are also summed to preserve spatial information. The final output layer returns the results. In our case, the results are represented by the segmented images, of the same size as the input ones, identifying the presence of diffuse radio emission.

A $3\times3$ pixels window (receptive window) is used for the convolution, applying it starting from the input images:
\begin{equation}
    s_{m,n}^f = \sum_{i,j=-1}^{1} w_{i,j}^f x_{m+i,n+j} + b^f,
\end{equation}
where $x_{m,n}$ is the $(m,n)$ pixel of the input image, $w_{i,j}^f$, and $b^f$ are the weights and the biases of the convolutional kernel. The result of the convolution, the $s_{m,n}^f$ element, constitutes the $f$-th {\it feature map}. Starting from the random initialization of the weights and the biases several different feature maps are created. Non-linearity is introduced by further processing the $s_{m,n}^f$ elements with an activation function. We have adopted the ReLU activation function \citep{Agarap18}:
\begin{equation}
    t_{m,n}^f = {\rm max}(0, s_{m,n}^f).
\end{equation}

In each level, the convolution plus activation steps are repeated two times. Batch normalization has been used after each convolution step to improve the convergence of the method. At this point, each feature map is downscaled through a max pooling function, which selects the maximum $t_{m,n}^f$ every 2$\times$2 pool of feature map elements. The resulting maps are 1/4 the original size. A dropout layer (with a rate of 0.5) is inserted after each max pooling function to prevent overfitting. The same convolution plus pooling procedure is repeated four times until the deepest layer is reached.  

Transpose convolutions with ReLU activation and upsampling layers are alternated in the expansive path, while dropout layers are not used. After five upsampling layers, specular to the contractive ones, the network produces 32 feature maps of the same size as the input image. A final convolution with softmax activation function provides two feature maps that represent, respectively, the probability of each pixel to be 0 or 1. The output is the probability map showing the probability of the $(m,n)$ pixel to be part of the diffuse radio emission:
\begin{equation}
    p_{m,n}^1 = \frac{e^{s_{m,n}^1}}{e^{s_{m,n}^0} + e^{s_{m,n}^1}}.
\end{equation}

During the training, weights and biases are optimized with an iterative procedure. Two sets of images are needed for the training: the input and the reference images. For the initial training of our network, we used as input synthetic observations of diffuse radio sources described in Sec.~\ref{sec:sinth}. The reference images are masks, displaying unity in pixels where emission from the diffuse radio source is present and zero otherwise. The input images are processed by the network and the segmented images are compared to the corresponding reference images using a loss function. We used the categorical cross-entropy loss function, defined as:
\begin{equation}
    H(y_{m,n},p_{m,n}) = -\sum_{f=1}^{N_c} y_{m,n}^f {\rm log}(p_{m,n}^f),
\end{equation}
where $y_{m,n}$ is the pixel value in the reference image (0 or 1) and $N_c$ is the total number of classes to classify. For our application $N_c=2$, that is, a pixel belongs or not to a diffuse radio source. The training aims at minimizing $H$ by updating the weights and biases with the backpropagation of the cross-entropy error estimate. This is done using the RMSprop algorithm as an optimizer. 

In addition to the trainable parameters (weights and biases), the network includes additional {\it hyperparameters} used to control the learning process. For our network, we selected the hyperparameters that demonstrated the best performance in \citep{Sanvitale2022}, ensuring their suitability for our purposes. They are:
\begin{itemize}
    \item the learning rate: $\mu=10^{-4}$, that regulates the step size of the iterative error minimization procedure;
    \item the batch size: $B=50$, that defines the number of images that are propagated through the network in one iteration step;
    \item the number of epochs: $E=200$ representing the number of times the training set is seen by the network during the training;
    \item the tile size: $T=192$, that defines the size of the input images (see Sec.~\ref{sec:trainandval}).
\end{itemize}
\smallskip

The network has been implemented using the Python programming language and exploiting the Keras software package \citep{chollet2015keras}, distributed as part of the Tensorflow framework \citep{tensorflow2015-whitepaper}, version 2.3.0. The resulting software can run on any computing platform, from standard multicore CPUs, available on a personal workstation, to supercomputers, with computational performance changing according to the adopted architecture. To speed up the computation, we have exploited the GPU implementation of Tensorflow.
\smallskip

Training and tests were first run on the Marconi100 (M100) HPC system\footnote{\url{https://www.hpc.cineca.it/hardware/marconi100}} and then moved on the new Leonardo supercomputer \footnote{\url{https://www.hpc.cineca.it/systems/hardware/leonardo/}} available at the CINECA Italian Supercomputing center. 

\section{LOFAR Observations}
\label{sec:lofar}

We developed Radio U-Net to detect diffuse radio sources in the LOFAR Two-meter Sky Survey \citep[LoTSS,][]{Shimwell17,Shimwell19}. This is an ongoing survey that aims to cover the entire northern sky with the LOFAR High-Band Antenna (HBA) at frequencies from 120 to 168 MHz (150 MHz central frequency). The second data release (DR2) of the LoTSS is fully described in \citet{Shimwell22}. The LoTSS DR2 covers 27$\%$ of the northern sky (5634 deg$^2$) and consists of images at 6$\arcsec$ resolution with a median root-mean-square (rms) sensitivity of 83 $\mu$Jy/beam. Released low-resolution images have 20$\arcsec$ restoring beam and median rms sensitivity of 95 $\mu$Jy/beam.

To test Radio U-Net on real data, we applied the network to a sample of galaxy clusters for which the search and classification of diffuse radio sources was made by tailored data reduction and visual inspection. These are the  309 clusters in the second catalog of Planck Sunyaev Zel’dovich \citep[PSZ2][]{Planck16b} that lie within the LoTSS-DR2 area (hereafter, the LoTSS-DR2/PSZ2 sample), presented in \citet{Botteon22}. We used their findings as a benchmark to test our approach (see Sec.~\ref{sec:results}). 

We note, however, that we did not use the images produced by \citet{Botteon22}. Instead, we used the public LoTSS DR2 data at 20$\arcsec$ resolution, directly downloaded from the archive\footnote{\url{https://www.lofar-surveys.org/dr2_release.html}}. Our aim is to test the segmentation and detection capability of Radio U-Net on basic archival data. If good performance is achieved, this approach will significantly reduce the computational time required for detection, which is typically performed on post-processed data.

\section{Synthetic Observations}
\label{sec:sinth}

\begin{figure*}
\begin{center}
\hspace*{-2cm}\includegraphics[width=1.2\textwidth]{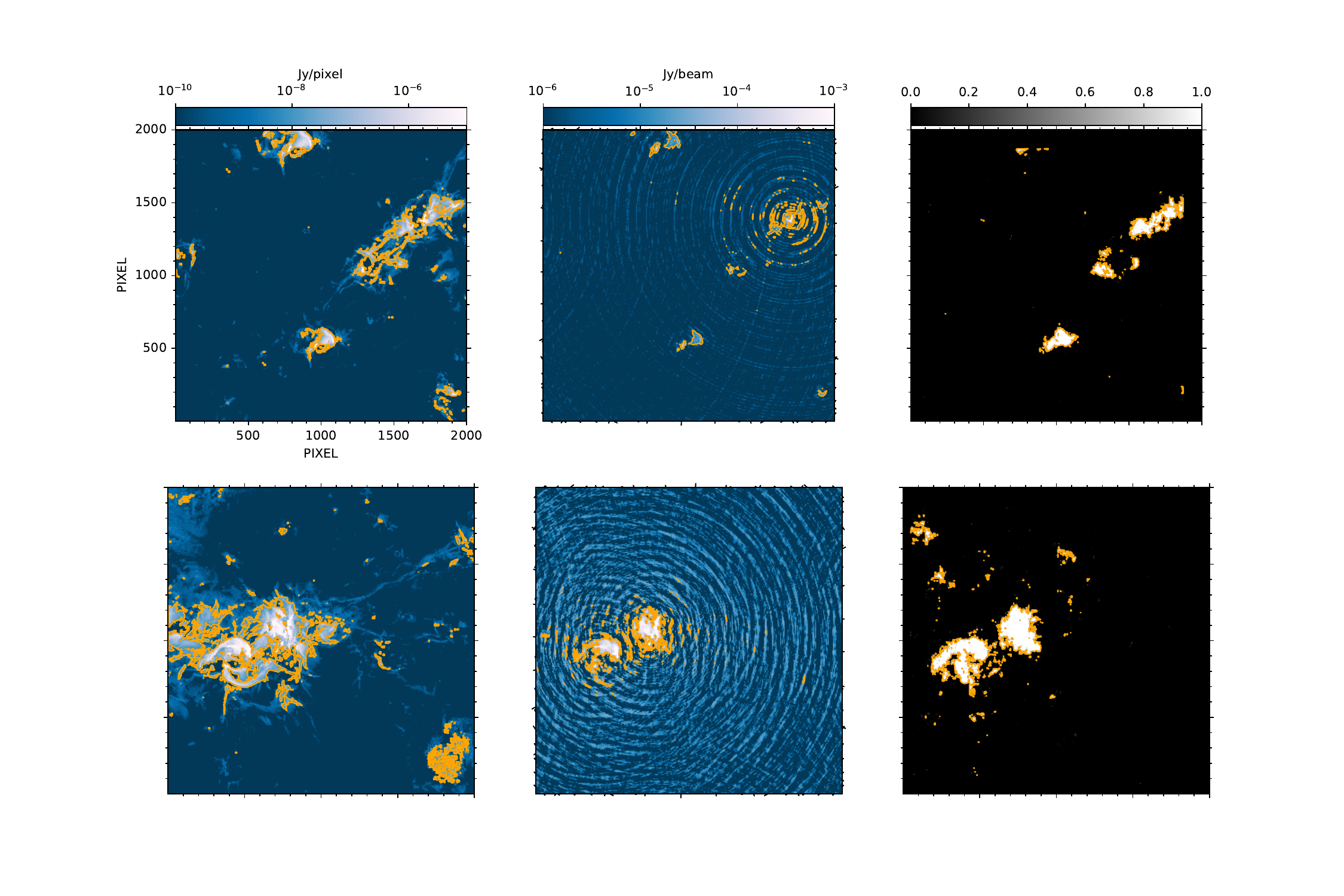}
\caption{Left panels: two examples of sky images with a contour level at $10^{-8}$~Jy/pixel, which represents the reference mask used to train the network. Central panels: clean images with contours drawn at 3$\sigma$, with $\sigma=2\times10^{-6}$~Jy/beam for the top image and $\sigma=1.5\times10^{-5}$~Jy/beam for the one at the bottom. The restoring beam is $6\arcsec$. Right panels: probability images created by Radio U-Net with contours at 0.5.}
\label{fig:synth}
\end{center}
\end{figure*}

We trained Radio U-Net on synthetic observations of diffuse radio emission. Input images have been generated to be the closest possible to actual LOFAR HBA observations, using the same procedure described in \citet{Gheller22}. The images are already publicly available (\url{https://owncloud.ia2.inaf.it/index.php/s/IbFPlCCcPUresrr}). Here we summarize the main steps of the procedure.

We used magneto-hydrodynamical (MHD) cosmological simulations, produced with the grid code {\it Enzo} \citep[][]{enzo14}. The simulation box has a uniform and constant spatial cell resolution of $41.65 \rm ~kpc$ (comoving) and covers a volume of $100^3 \rm ~Mpc^3$ simulated with $2400^3$ dark matter particles and cells. A uniform primordial magnetic seed field of $B_0=0.1 ~\rm nG$ (comoving) was initialized in all directions at $z_{\rm in}= 45$, and was evolved assuming ideal MHD, via adiabiatic compression/rarefaction and small-scale dynamo amplification in halos, until $z=0$.
{\bf The simulation assumed a standard $\Lambda$CDM cosmological model, with
density parameters $\Omega_{b} = 0.0478$, 
$\Omega_{\rm DM} = 0.2602$,  $\Omega_\Lambda = 0.692$, and a Hubble parameter $H_0=67.8 ~\rm km/s/Mpc$.}

We computed the emission at 150 MHz from relativistic electrons accelerated by cosmic shocks at various redshifts,  assuming that only shocks can accelerate relativistic particles via diffusive shock acceleration, and produce synchrotron radio emission\citep[e.g.][]{2011JApA...32..577B}. For simplicity, the synchrotron emission model by \citet{hb07} has been assumed here, which requires the jump condition of each cell undergoing shocks (computed from the simulation), the local value of the magnetic field, and the electron acceleration efficiency as a function of Mach number (which is calibrated on shocks internal to galaxy clusters, as in \citealt{va15radio} and \citealt{va19}). We did not include in our simulations radio emission produced by turbulence, as is expected from radio halos, bridges, or mega halos \citep{Brunetti11b,Brunetti20,Nishiwaki24}, as well as by electrons re-accelerated by shocks. However, the morphology and emissivity of diffuse radio sources can loosely resemble even those of radio halos, despite the different particle acceleration mechanisms likely at work in radio halos \citep[e.g.,][]{vanWeeren19,Lee24}.
This same simulation and model for radio emission was successfully used to compare with the result of stacking attempts of the cosmic web with real radio data \citep{vern20,2023SciA....9E7233V}, and therefore it can be considered as a fairly realistic representation of how the radio emitting cosmic web might look like. 

The production of the training set of synthetic images for Radio U-Net, starting from the 3-dimensional emission model of the cosmological simulations, followed two main basic steps.

First, we integrated the emission along the line of sight within four comoving volumes chosen at four different redshift snapshots
(roughly equally spaced from 
$z=0.02$ to $z=0.15$). An example of the maps of some of the redshift snapshots used is given in Fig.~2 in \citet{2018MNRAS.480.3749G}. We applied cosmological corrections for the surface brightness and the luminosity distance
of each redshift, relative to the $z=0$ observer. We also decreased the pixel size of each image as a function of distance, using a cubic interpolation on the input map. Artifacts arising from the periodicity of structures along the same line of sight were minimized by shifting each box with a random offset in both directions. 

Next, every single synthetic sky model was generated by progressively stacking maps referred to an increasing redshift. In order to fully cover the $\approx 640 ~\rm Mpc$ distance out to $z=0.15$, we extracted maps multiple times from the same simulated redshifts, applying an appropriate cosmological corrections factor to simulate an increasing redshift of the sources, similar to \citet{2021MNRAS.500.5350V}. As done in \citet{Gheller22}, we limited our analysis of simulated lightcones up to $z \approx 0.15$ since we do not expect a drastic change in the radio emission including larger redshifts \citep[see also][]{2021PASA...38...47H}.

This procedure was repeated many times, by applying random rotations to each of the different redshift slices, so that we could obtain more than 500 independent 
 lightcones. The images have a resolution of $2000 \times 2000$ pixels, sampling a field of view of $1.1^\circ\times1.1^\circ$ each, with a nominal angular resolution (pixel size) of 2 arcsec. These images are indicated as {\it sky images} (see Fig.~\ref{fig:synth}, first column, for two examples). 
 
 The reference images used to train the network are created by masking the sky images at $10^{-8}$~Jy/pixel. This value is a factor $\sim10^{-3}$ lower than the nominal noise level (83~$\mu$Jy/beam = $2\times10^{-6}$~Jy/arcsec$^2$ with a beam size of 6$\arcsec$). This is done to teach the network to recognize emission structures even below the noise level.

To generate mock observations, the synthetic visibilities corresponding to a LOFAR HBA observation of 8 hours have been calculated using the ``predict'' mode of the WSClean with the sky images as model images \citep{offringa-wsclean-2014,offringa-wsclean-2017}. At this point, random noise is added to the visibilities using the noise.py script in the LoSiTo software package\footnote{\url{https://github.com/darafferty/losito/tree/master}}. Finally, imaging is performed using the WSClean software, adopting Briggs' weighting, with ‘‘robust’’ parameter equal to 0, and correcting for the primary beam. The use of these parameters resulted in a restoring beam of $5.9\arcsec\times5.1\arcsec$. Deconvolution is carried out using the clean method and the auto-masking option with a 5$\sigma$ threshold.

We note that the cleaning process has not been optimized for each image, as it was automatically applied to hundreds of images. Residual artifacts influence the quality of the final image and can potentially affect any attempt at automating source identification, as these artifacts may be confused with real sources. This is an intended behavior designed to train the network to recognize sources in imperfectly cleaned images. The random noise previously introduced in the visibilities produces a rms of $\sim2.3\times10^{-7}$~Jy/beam in source-free fields, while the median rms is about one order of magnitude higher due to residual imaging artifacts. Restored images are used as input for the training of Radio U-Net, and we will refer to these images as {\it clean images} (see Fig.~\ref{fig:synth}, central column, for two examples).

\section{Training Strategy and Validation of the Network}
\label{sec:trainandval}

Training and validation of Radio U-Net were performed on synthetic observations. The primary objective of this training is to enable the network to identify a wide range of cosmological source morphologies and to distinguish real sources from imaging artifacts. Additionally, the reference masks used in the training were created with a threshold lower than the actual noise level of the input clean images, allowing the network to recognize faint patterns even below the noise.

The $2000 \times 2000$ pixels images are further divided into square tiles that become the actual training set of the network. The tile size is chosen to be the smallest possible, still representative of the features to be identified (both sources and artifacts, in particular, those due to the dirty beam, which can span large areas of the image). In this way we maximize the size of the training sets without losing the significance of each single input image, reducing, at the same time, the memory footprint of the training. An effective tile size for our images results to be 192$\times$192 pixels so that from each image we can create 100 tiles.

Due to GPU memory constraints,  the training set comprises 100 sky images, i.e 10000 tiles. A fraction of them ($5\%$, that is, 500 tiles) is kept as a validation set to verify the performance of the network during training. Another 10 randomly selected images (i.e. 100 tiles), never seen by the network, are used as a test set.

The workflow is implemented as follows:
\begin{enumerate}
\item the training program reads the input parameters, and the hyperparameters and sets up the network;
\item training images (input clean images and reference masked images) are read from FITS files, and stored on disk;
\item the data are transformed in logarithmic scale and normalized in a [0,1] range, using $10^{-10}$~Jy/pixel and $10^{-5}$~Jy/pixel as minimum and maximum data normalization values. Pixels outside this range were clipped to the boundaries of the normalization range;
\item images are divided into tiles of 192$\times$192 pixels;
\item tiles are serialized to feed the network;
\item mini-batches of tiles are offloaded to the GPU and processed for the training;
\item at each epoch, the loss function value is computed for the training and the validation sample;
\item the trained network is saved in a file (in ‘‘SavedModel’’ Tensorflow format).
\end{enumerate}

\begin{figure}
\begin{center}
\hspace{-0.9cm}\includegraphics[width=1.1\linewidth]{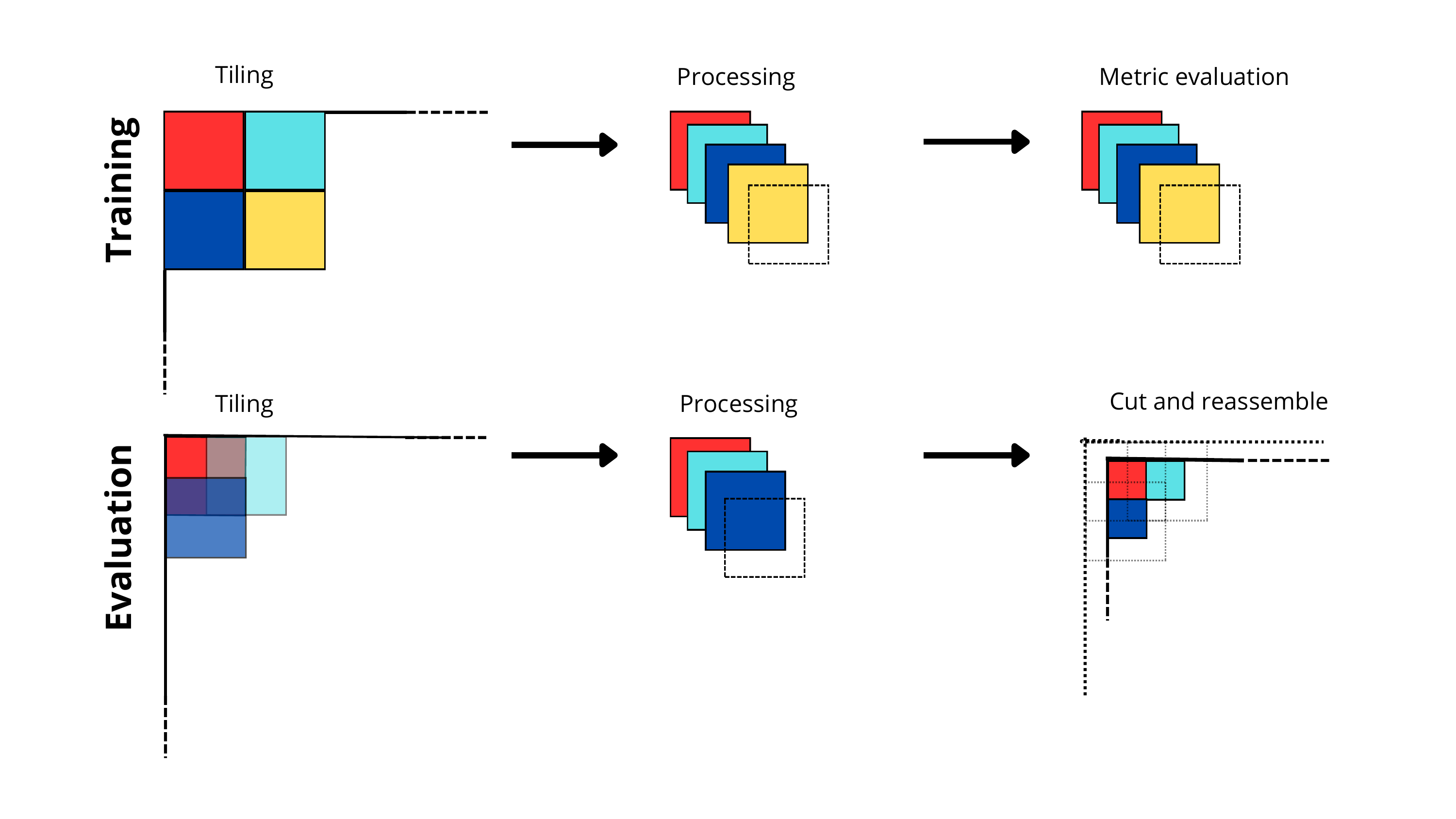}
\caption{Scheme of the tiling procedure. During training, the tiles are all independently processed. The categorical cross-entropy loss function, which is the metric used during the training, is computed for each tile separately. When the evaluation program is used to apply the trained network to new images, the tiles are overlapped. They are still processed independently and the central cut-outs are reassembled to avoid boundary problems. The size of the figures can vary but the tile size is fixed to 192 pixels.}
\label{fig:tiling}
\end{center}
\end{figure}

Weights and biases of the trained network can be loaded by the evaluation program which performs the segmentation of new images and saves the result into a FITS file. The only significant limitation of the tiling procedure is the potential for minor misalignments at tile borders, which may compromise the accuracy of segmentation in the reconstructed images. To mitigate this issue, when Radio U-Net is used, tiles are created with overlapping boundaries, which are then discarded in the reconstruction. The tiles are overlapped by half their size (96 pixels) and only the central 48x48 pixels of each tile are then used to recompose the final image. An explicative scheme of this procedure is shown in Fig.~\ref{fig:tiling}. Residual boundary effects are still visible in the probability image but they do not affect the detection of the diffuse emission.

Examples of the resulting segmented images from the test set are shown in Fig.~\ref{fig:synth}, third column. This probability image has to be compared with the sky image, in the first column, where the contours show the reference mask on which the network was trained, and with the input clean image, shown in the second column. In all 10 images, the diffuse emission is spotted against noise and cleaning artifacts, showing high detection probability values. The broad structure of the sources is well recovered, and, although the faintest features are missed, it is important to remember that these features are orders of magnitude below the actual noise in the input images. For comparison, we plotted the 3$\sigma$ contours over the clean images (with $\sigma$ computed as the rms in an empty region of the image). These contours delineate the regions that would be detected by a standard analysis. The striking result is that the segmented images produced by Radio U-Net enable much easier identification of diffuse radio structures, reproducing their complex morphologies.

A binary mask can be generated from the segmented image by setting a pixel probability threshold. We notice that the boundaries between the detected and undetected regions are very sharp, and a threshold of 0.5 well bounds all the features detected by the network (see Fig.~\ref{fig:synth}). By comparing the reference mask and the segmented image, we can estimate how many pixels are true (true 0, T0, and true 1, T1) or false identification (F0 and F1) using 0.5 as a threshold.

We estimate for each image the mean intersection over union (IoU):
\begin{equation}
    {\rm IoU = 0.5\Bigl(\frac{T0}{T0+F0+F1}+\frac{T1}{T1+F1+F0}\Bigr)} ~.
\end{equation}

This is the benchmark evaluation metric for computer vision tasks for which the correct identification of pixels with value 1 is as important as that of pixels with value 0: it is equal to 1 in case of perfect segmentation and decreases for increasing misidentified pixels. This formula simplifies to a single-class metric when only one class is present in the image. Averaging the 10 test images, we obtain an IoU of 0.64$\pm$0.02. This number indicates a good segmentation, considering that the reference mask is orders of magnitude below the noise of the input images.

If we compute the same metric using masks created from clean images at the standard $3\sigma$ level (e.g., the orange contours in the second column of Fig.~\ref{fig:synth}) and compare with the reference mask, we obtain an IoU of 0.58$\pm$0.02, averaged over the 10 images. This demonstrates that the network is able to produce a more reliable segmentation with respect to the standard method.

Furthermore, while the 3$\sigma$ threshold is used for detection only on fully cleaned images, for which tailored data reduction and imaging procedures are often performed, Radio U-Net segmentation performs well even with incompletely cleaned images. Once trained, the method does not require parameter tuning, other than fixing the probability threshold for detection. It performs the segmentation of each single 2000$\times$2000 image in 0.7 s, faster than any procedure requiring human interaction. This feature decreases the amount of time required to reduce data for detection, making it suitable for use in large radio surveys to conduct blind searches for diffuse radio emissions. 

However, we can exploit multiwavelength analyses to assist in identifying diffused radio sources associated with galaxy clusters, bridges, and cosmic web filaments.
In this case, we can apply Radio U-Net to a list of targets and classify them as detected or not on the basis of the segmentation mask created by the network. This is the procedure that we will follow in the next Section, where we will also compute precision and recall for binary classification on a test sample of galaxy clusters. 

In binary classification problems, precision is defined as the ratio of the true positive (TP, detected diffuse radio sources) to the total number of objects identified as detected (both true and false positive, FP), and it reaches unity if there are no false positives:

\begin{equation}
\label{eq:precision}
    {\rm precision} = \frac{TP}{TP+FP} ~.
\end{equation}

Recall is the ratio of the TP to the sum of the true positive and false negative (FN):

\begin{equation}
\label{eq:recall}
    {\rm recall} = \frac{TP}{TP+FN} ~.
\end{equation}

We can consider as TP all the sources of the sky images detected with more than 10 pixels (i.e. the size of the resolution element of the synthetic observation) above 0.5 in the probability images. In contrast, FP are the regions above the 0.5 level in the probability images but without any counterpart in the sky images, whereas FN are all the sources that remain undetected. We did not consider the sources outside the boundary which is cut by the tiling procedure. 

In the set of 10 sky images used for the test, we achieved a precision of 0.86 and a recall of 0.65. It is noteworthy that $35\%$ of the TP can only be detected in the images segmented by Radio U-Net, as they are not visible in the clean images above the 3$\sigma$ threshold.

\section{Results}
\label{sec:results}

Having trained Radio U-Net on a large sample of synthetic observations, we can now use it to process real LOFAR observations. We will use as a reference the LoTSS-DR2/PSZ2 cluster sample, introduced in Sec.~\ref{sec:lofar}, to test the classification performances of the network.

\subsection{Application to the LoTSS-DR2/PSZ2 cluster sample}

Low-resolution (20$\arcsec$) images of the 309 clusters in the sample were downloaded from the LoTSS archive. For each cluster, we created a cut-out of 960$\times$960 pixels centered on the cluster position, ensuring the target was as close as possible to the pointing center of the mosaic. The cut-out size was selected to be as large as possible, ensuring that it does not exceed too much the boundary of the LOFAR primary beam, and it is also a multiple of 192, which remains the tile size for the network. 

The pixel size is 4.5$\arcsec$, hence the cutout size is $1.2^\circ\times1.2^\circ$. Redshift, available for all but 28 clusters, spans between 0.016 and 0.9, with a median value of 0.28 \citep{Botteon22}. By adopting the same cosmology model (i.e., $\Lambda$ cold dark matter, with $\Omega_{\Lambda} = 0.7$, $\Omega_{m} = 0.3$, and $H_0 = 70$~km~s$^{-1}$ ~Mpc$^{-1}$), 1~Mpc corresponds to 682 pixels at $z = 0.016$ and to 28 pixels at $z = 0.9$. This means that, if present, cluster-scale radio emission should be contained in the cut-out and it can span a broad range of angular extensions.

We processed the 309 images with Radio U-Net using a slightly modified version of the evaluation program which allows us to preserve the astrometric information contained in the header. As in step $(iii)$ of the training procedure described in Sec.~\ref{sec:trainandval}, we converted the data in logarithmic scale and we normalized between 10$^{-7}$ and 10$^{-2}$ Jy/beam. We thus obtained probability images for all the galaxy clusters.

\begin{figure*}
\begin{center}
\hspace*{-2cm}\includegraphics[width=1.2\textwidth]{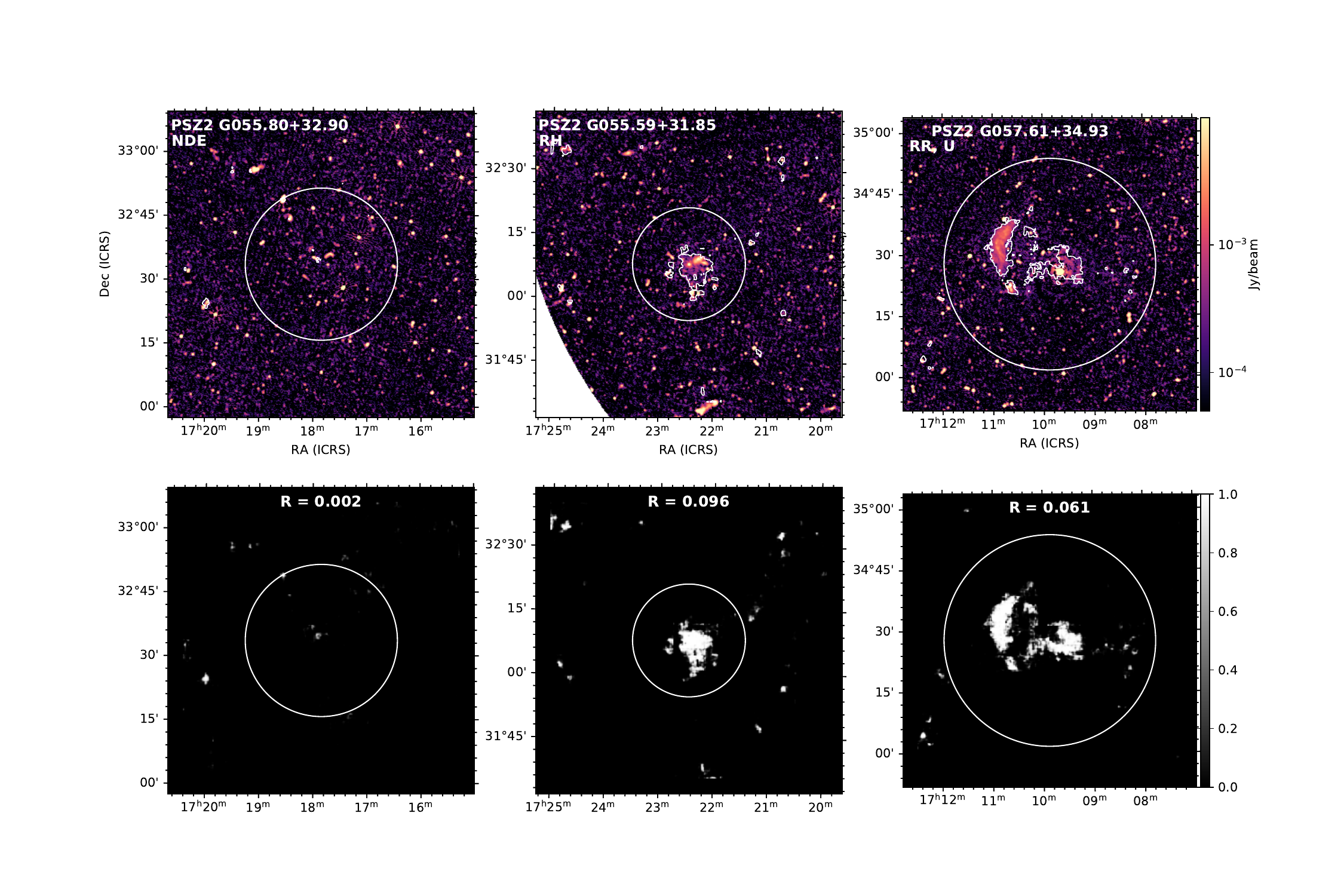}
\caption{Examples of LoTSS-DR2/PSZ2 galaxy clusters processed by Radio U-Net and successfully classified on the basis of their $\mathcal{R}$-value (see Sec.~\ref{sec:results}). Low-resolution ($20\arcsec$ beam) images of the LoTSS survey are shown in the first row while their probability maps are shown in the bottom row. The 0.5 probability contour is also reported in the images of the first row. The white circle shows the reference 2.2$R_{500}$ radius where $\mathcal{R}$ is computed. NDE marks a non-detection, RH and RR mark the presence of a radio halo and a radio relic, respectively. U is used to classify an uncertain detection that, in this case, is a radio-halo.}
\label{fig:good_example_orig}
\end{center}
\end{figure*}

\begin{figure*}
\begin{center}
\hspace*{-2cm}\includegraphics[width=1.2\textwidth]{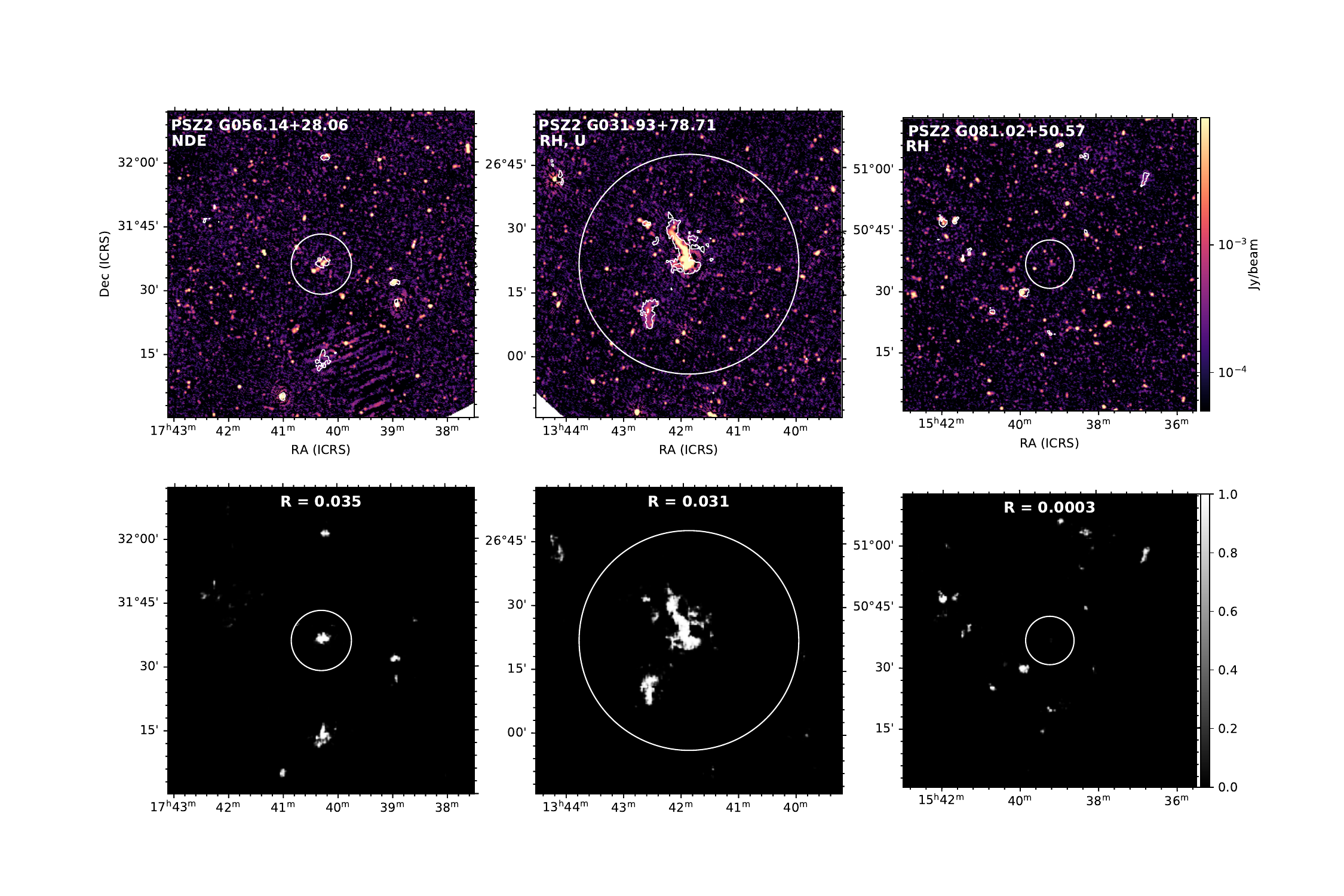}
\caption{Examples of LoTSS-DR2/PSZ2 galaxy clusters processed by Radio U-Net (see Sec.~\ref{sec:results}). Low-resolution ($20\arcsec$ beam) images of the LoTSS survey are shown in the first row while their segmented maps are shown in the bottom row. The white circle shows the reference 2.2$R_{500}$ radius where $\mathcal{R}$ is computed. The 0.5 probability contour is also reported in the images of the first row. NDE marks a non-detection, RH marks the presence of a radio halo while U is used to classify an uncertain detection. The first column shows a cluster incorrectly classified as detected. The second column shows a case where the cluster is correctly classified as detected but the probability map is dominated by the central extended radio galaxy rather than by the radio halo. The third column shows a case of a radio halo not detected by Radio U-Net. }
\label{fig:bad_example_orig}
\end{center}
\end{figure*}

We show some examples of LoTSS-DR2/PSZ2 clusters with the corresponding output generated by Radio U-Net in Fig.~\ref{fig:good_example_orig} and Fig.~\ref{fig:bad_example_orig}. In particular, Fig.~\ref{fig:good_example_orig} shows examples of clusters where the direct application of Radio U-Net gave very good segmentation. PSZ2 G055.80+32.90 is a galaxy cluster with no signs of diffuse radio emission; indeed, very few pixels show a probability greater than 0.5 of being part of a diffuse source. In contrast, PSZ2 G055.59+31.85 hosts a radio halo, and its shape is clearly recovered by the network. Notably, in many cases, although Radio U-Net is applied to images with the contaminant emission from compact sources and extended radio galaxies, it successfully recovers the morphology of the emission that was only identifiable in lower-resolution source-subtracted images in \citet{Botteon22}. PSZ2 G057.61+34.93 hosts a radio relic \citep[a.k.a. the \textit{Cornetto} relic,][]{Locatelli20} and a candidate radio halo, which was classified as uncertain due to the presence of a bright radio source that generates residual artifacts in the halo region. Since Radio U-Net was trained to uncover diffuse radio emission even beneath residual imaging artifacts, the probability map obtained for this cluster allows us to confirm the presence of the central radio halo.

Fig.~\ref{fig:bad_example_orig} shows examples of poorly segmented images. PSZ2 G056.14+28.06, which does not host diffuse radio emission according to \citet{Botteon22}, appears to host a central radio halo in the Radio U-Net probability map. On the contrary, PSZ2 G081.02+50.57, which hosts a radio halo visible only in lower-resolution images, is completely missed by Radio U-Net. Additionally, the bright and extended radio galaxy at the center of PSZ2 G031.93+78.71 dominates the probability map of this cluster, obscuring the emission of the halo. The fact that Radio U-Net also detects extended radio galaxies indicates its ability to generalize and suggests its potential use in searching for such objects. However, this also highlights the need to develop better strategies to distinguish between different types of extended radio emissions.

\subsection{Classification accuracy}

\begin{table}
    \centering
    \begin{tabular}{c|c|c|c|c}
      &  NDE & DE & DEU & tot \\
    \hline
    Initial test set &  114 & 85 & 47 & 246 \\
    \hline
    Fine-tuning set & 0 & 78 & 0 & 78 \\
    \hline
    Test set & 47 & 0 & 47 & 94 \\
    \end{tabular}
    \caption{Cluster sub-samples. We list the number of non-detection (NDE), detection (DE), detection with uncertain classification (DEU) contained in the initial test set, the fine-tuning set, and the test set, respectively.}
    \label{tab:samples}
\end{table}

\citet{Botteon22} classified the radio sources detected in each cluster as radio halo (RH), radio relic (RR), candidate radio halo or relic (cRH, cRR, if no X-ray information was available for the cluster), and uncertain (U) when the image was significantly affected by calibration artifacts or the radio emission did not show a morphology, size, and/or position consistent with the previous categories. If no diffuse emission was detected, the cluster is classified as NDE, while in case of poor data quality, the classification is not applicable (N/A). Hereafter, we will use the same classification scheme, but we will also use the umbrella term diffuse emission (DE) to encompass RH, RR, cRH, and cRR, since we are not interested in distinguishing different kinds of cluster diffuse emission at the moment. Clusters with both DE and/or uncertain sources are classified as DEU. 

For testing purposes, we selected all NDE (114), DE (85), and DEU (47) clusters having redshift estimates. This will be our initial test set comprising a total of 246 galaxy clusters (see Tab.~\ref{tab:samples}). This sub-sample is well balanced, having similar numbers of positive (DE and DEU) and negative (NDE) instances. To claim a detection, i.e. classify a cluster as positive or negative, we considered the sum probability of all the pixels within the cluster area which we defined as a circle with radius 2.2$R_{500}$ \citep[where $R_{500}$ for each cluster is provided by][]{Botteon22}. This is the maximum distance at which radio relics were detected in this sample \citep{Jones23}. We define $\mathcal{R}$ as the ratio of the sum probability and the total number of pixels (N) within the circle:

\begin{equation}
    \mathcal{R} = \frac{\sum_{m,n} p_{m,n}^1}{N}.
\end{equation}

$\mathcal{R}$ is a proxy for the extension of the source within the cluster: it approaches 0 when no source is detected while it is 1 when a diffuse source occupies the entire circle and has all pixels detected with probability 1.

\begin{figure}
\begin{center}
\includegraphics[width=\linewidth]{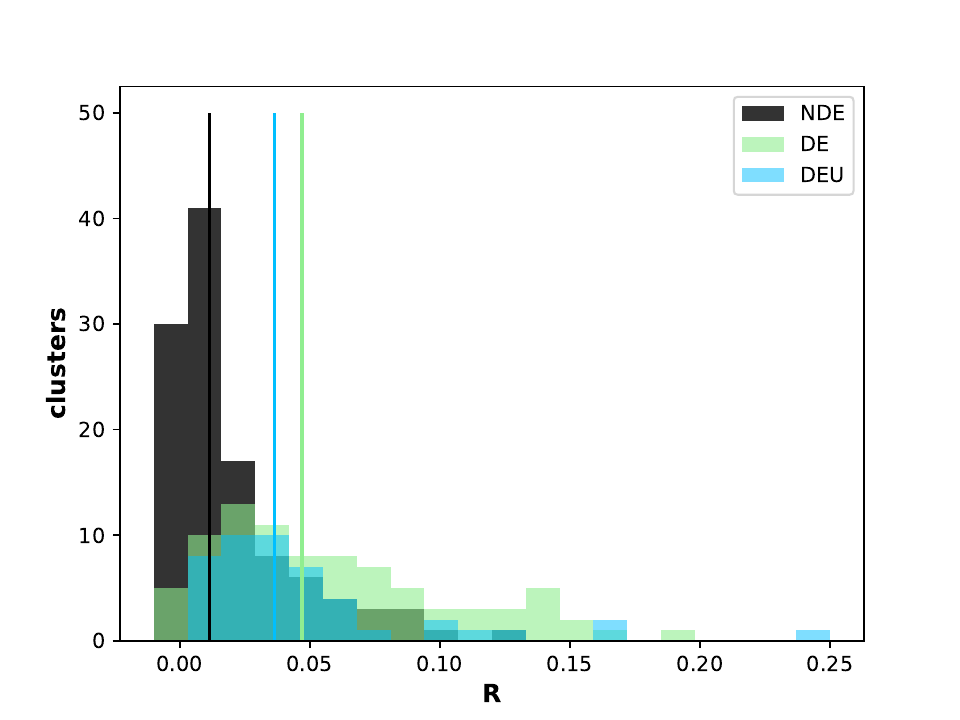}
\includegraphics[width=\linewidth]{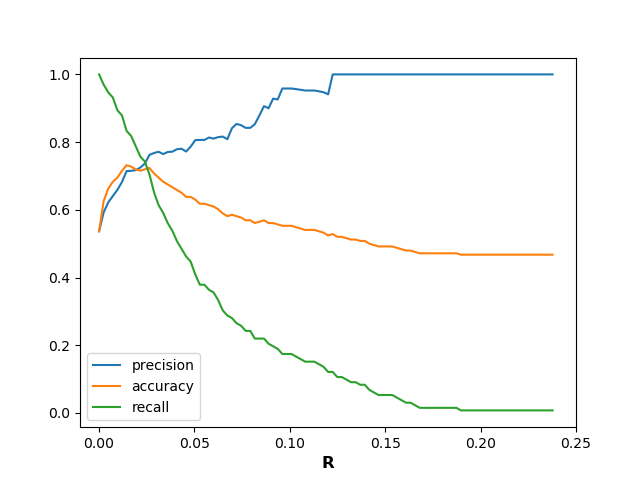}
\caption{Top panel: distribution of $\mathcal{R}$ for non-detected (NDE) clusters and for detected diffuse emission (DE) and uncertain detection (DEU) clusters, with a vertical line showing the median of the distribution. Bottom panel: Precision, accuracy, and recall for the binary classification between false (NDE) and positive (DE and DEU) computed in the initial test set for different values of detection threshold on $\mathcal{R}$.}
\label{fig:stat_orig}
\end{center}
\end{figure}

In Fig.~\ref{fig:stat_orig} we show the distribution of $\mathcal{R}$ for NDE, DE, and DEU clusters and their median values. Clearly, the distributions are different for clusters with and without detections, with the peak of the distribution being lower for NDE clusters. We note that the distribution of DE and DEU clusters is instead similar and they can be drawn from the same distribution according to the two-sample Kolmogorov-Smirnov (KS) test with a confidence level of $95\%$. This indicates that Radio U-Net recognizes diffuse radio emission regardless of the specific classification. However, it is not straightforward to define a threshold value for $\mathcal{R}$ to separate NDE from DE and DEU clusters.

In binary classification problems, each object belongs to a class and is labeled as positive or negative. The model assigns a predicted label to each object, which can be true if the classification is correct or false if it is incorrect. In our case, the predicted label depends on the threshold set on $\mathcal{R}$ to distinguish between galaxy clusters with detected (positive) and non-detected (negative) diffuse radio sources. Consequently, the total number of true classifications (TP and true negative, TN) and false classifications (FP and FN) also depends on $\mathcal{R}$. To choose the optimal threshold for $\mathcal{R}$, we calculated various standard performance metrics as a function of $\mathcal{R}$.

The accuracy is defined as the fraction of correct classifications to the total number of objects:

\begin{equation}
    {\rm accuracy} = \frac{TP+TN}{TP+TN+FP+FN} ~.
\end{equation}

This is an important metric but can be misleading in the case of unbalanced datasets and it weighs equally between false positive and false negative. For this reason, we also computed precision and recall (Eqs.~\ref{eq:precision} and \ref{eq:recall}). While precision highlights the purity of positive predictions, recall focuses on the network's ability to capture all positive instances. Therefore, precision can be high even if the model misses some positive object, while the recall will be low. For example, using Radio U-Net to make a first selection of interesting objects in an all-sky survey, we may be more interested in obtaining a high recall since false positive detections could be discarded subsequently by a more tailored examination of the images. Conversely, if we are interested in selecting a small, pure sample of galaxy clusters hosting diffuse radio emissions, we would prefer high precision.

The accuracy, precision, and recall obtained on the initial test sample as a function of the threshold imposed to $\mathcal{R}$ are shown in the bottom panel of Fig.~\ref{fig:stat_orig}. The accuracy has a maximum at $\mathcal{R}$=0.015 where it reaches $73\%$. Since our dataset is almost balanced, this means high accuracy both in the positive and in the negative classifications. At $\mathcal{R}$=0.015 the model also reaches a high recall ($83\%$), therefore, we decided to use this value as a reference. The precision, recall, and accuracy obtained for the initial test sample using the model trained on synthetic observations (i.e., the original model) with $\mathcal{R}$=0.015 as threshold are listed in Tab.~\ref{tab:results}.

These values cannot be directly compared with the precision and recall that we derived for synthetic observations in Sec.~\ref{sec:trainandval}. This discrepancy arises because positive instances are here defined based on the analysis of actual observations, while for synthetic observations they are based on the sky images that reach below the noise level. This can explain the lower recall obtained for synthetic observations (i.e. 0.65). Additionally, our simulations do not include radio galaxies and star-forming galaxies; hence, false positives in synthetic observations are solely due to imaging artifacts. This explains the higher precision obtained for synthetic observations (i.e. 0.86).

Using $\mathcal{R}$=0.015 as threshold we obtain 71 clusters correctly identified as NDE and 109 correctly detected, among which 70 are DE and 39 are DEU. Within the 23 missed detections, 15 are DE and 8 are DEU, implying that in both sub-classes, around $17\%$ of the detections are missed. In Sec.~\ref{sec:discussion} we will discuss in more detail the properties of the clusters which are incorrectly identified.

\begin{table}
    \centering
    \begin{tabular}{c|c|c|c}
     & precision & recall & accuracy \\
    \hline
    Initial test set & & & \\
    Original$_{\mathcal{R}=0.015}$ & 0.72 & 0.83 & 0.73 \\
    \hline
    Test set & & & \\
    Original$_{\mathcal{R}=0.026}$ & 0.76 & 0.72 & 0.70 \\
    FT$_{\mathcal{R}=0.00084}$ & 0.70 & 0.89 & 0.73 \\
    FT-DA$_{\mathcal{R}=0.00081}$ & 0.70 & 0.87 & 0.72 \\
    \end{tabular}
    \caption{Precision, recall, and accuracy obtained. These values are shown for the initial test set and the test set (see Tab.~\ref{tab:samples}) processed with the original model (original), the model retrained with fine-tuning (FT) and the model re-trained using fine-tuning and data augmentation (FT-DA). The $\mathcal{R}$ values used as a threshold to compute the metric are also shown.}
    \label{tab:results}
\end{table}

\subsection{Fine-tuning and data augmentation}
\label{finetuning}

Considering that Radio U-Net was trained exclusively on synthetic observations, achieving a detection accuracy of $73\%$ is a significant result. However, to enhance the network's performance, we have explored transfer learning using LoTSS observations. This procedure, known as fine-tuning, can be applied to various types of neural networks, including U-Net \citep{Amiri20}. A widely used approach involves transferring the weights directly from the pre-trained model and continuing the training with new data. This approach typically optimizes the weights of only the last layers while freezing the other layers. The number of layers that has to be re-trained depends on the sample size and on the similarity between the original and new datasets.

The network architecture consists of 76 layers. Using the fine-tuning technique, we can freeze most layers of the model, whose weights are trained on simulated data, and re-train only the uppermost layers of the expansive path using the LoTSS dataset. The training process tunes the weights to features associated specifically with the LoTSS-DR2/PSZ2 dataset.

To implement fine-tuning, we need the reference masks for some of the LoTSS-DR2/PSZ2 galaxy clusters. We created these masks starting from source-subtracted images generated with a Gaussian taper of $15\arcsec$ \citep[see][for more details on the subtraction procedure]{Botteon22}. We used \texttt{breizorro}\footnote{\url{https://pypi.org/project/breizorro}} to automatically create masks with a $5\sigma$ threshold. Additionally, we manually inspected the masks to exclude sources unrelated to the diffuse emission detected in the cluster and residual artifacts. This step was necessary because the subtraction procedure, which was automatically applied to all targets, may have left residual diffuse emission from extended radio galaxies in the image.

For the training we only used DE clusters, therefore excluding clusters with some kind of uncertain detection which will be instead used for the test. Therefore, the training set comprises 78 galaxy clusters, while for the test we used 47 clusters with detection, and an equal number of NDE clusters to have a balanced dataset (see Tab.~\ref{tab:samples} for a summary). The training set was further divided into $90\% - 10\%$ for validation purposes.

It is important to note that, to ensure the convergence of the training, the ratio of the number of tiles that contain no signal to that of tiles containing some signal must be kept in the order of unity. This is achieved first by reducing the size of the images to $576\times576$ pixels (that is, 9 tiles in each image). However, for most of the images, the diffuse emission only occupies only 1 or 2 tiles out of nine (for a total of 125 tiles out of 702). Hence, we decided to try two different strategies, the first consisting of discarding some of the empty images and the second increasing the number of tiles with data augmentation.

\begin{figure}
\begin{center}
\includegraphics[width=\linewidth]{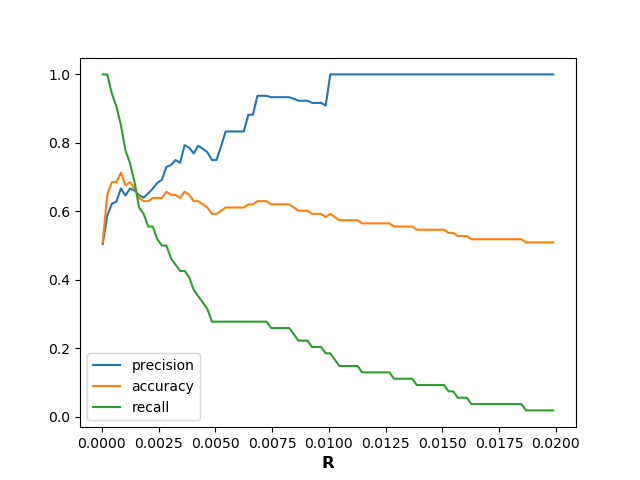}
\caption{Precision, accuracy, and recall of the fine-tuned network computed in the test set for different values of detection threshold on $\mathcal{R}$.}
\label{fig:stat_retrained}
\end{center}
\end{figure}

Using the first strategy, we obtain a total of 250 tiles for the transfer learning process. We tested different fine-tuning strategies, varying hyperparameters such as the number of epochs and the learning rate, and changing the number of trained layers. We obtained the best improvement on the training and validation sample with the fine-tuning of the uppermost 10 layers with 100 epochs and a learning rate of $10^{-4}$. However, when applied to the test sample for image classification, the fine-tuned network does not perform better than the original one, giving a similar value of maximum accuracy ($73\%$) on the test set, for $\mathcal{R}=8.4\times10^{-4}$ (see Fig.~\ref{fig:stat_retrained} and Tab.~\ref{tab:results}). We note that the $\mathcal{R}$ values obtained with the fine-tuned model are one order of magnitude lower than with the original model, which means that in general a lower number of pixels is detected. The recall slightly improves, with only $11\%$ of the diffuse emission hosting clusters missed. However, this is compensated by the lower precision value reached at the maximum accuracy (see Fig.~\ref{fig:stat_retrained} and Tab.~\ref{tab:results}).

Considering the limited number of images available for fine-tuning, we used data augmentation to increase the variety of the training sample. We apply rotations of 90, 180, and 270 degrees and vertical and horizontal flipping to the 78 images of the training set. This procedure increases the number of tiles to 1500. However, we only obtained again a similar maximum accuracy reached at $\mathcal{R}=8.1\times10^{-4}$  ($72\%$, see also Table~\ref{tab:results}) 

We thus conclude that our observational dataset is not big and rich enough to be used for the fine-tuning of the network. Adopted data augmentation schemes, using rotation and reflection of training images, are not sufficient to increase the complexity of the training sample and cannot improve the completeness of the dataset lacking extensive information on the complex and diverse morphologies of the investigated sources. However, other augmentation strategies, such as zooming or brightness and contrast adjustment, are not suitable for our case because they would change the resolution beam and flux scale of the image. 

\section{Discussion}
\label{sec:discussion}

\begin{figure*}
\begin{center}
\hspace*{-2cm}\includegraphics[width=1.2\textwidth]{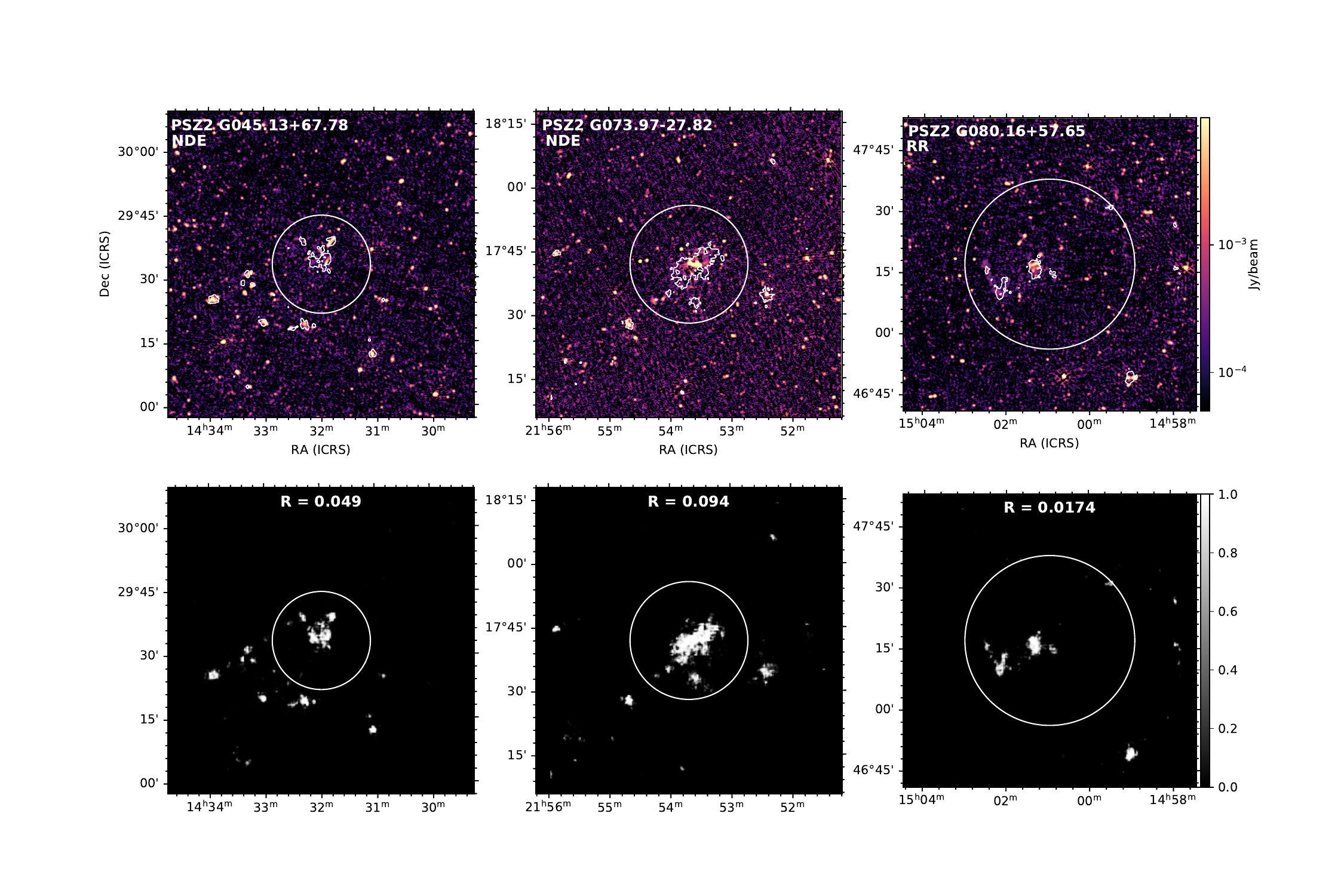}
\caption{Examples of LoTSS-DR2/PSZ2 galaxy clusters processed by Radio U-Net with false positive detections (see Sec.~\ref{sec:discussion}). Low-resolution ($20\arcsec$ beam) images of the LoTSS survey are shown in the first row while their segmented maps are shown in the bottom row. The 0.5 probability contour is also reported in the images of the first row.  The white circle shows the reference 2.2$R_{500}$ radius where $\mathcal{R}$ is computed. NDE marks a non-detection according to \citet{Botteon22}, while RR marks the presence of a radio relic. The left and central panels show two examples where Radio U-Net detected the presence of diffuse radio emission not only associated with radio galaxies which was not detected in \citet{Botteon22}. In the right panel another similar example in which the network spotted the presence of the radio halo which was previously detected in \citet{vanWeeren21}.}
\label{fig:fp}
\end{center}
\end{figure*}

\begin{figure*}
\begin{center}
\hspace*{-2cm}\includegraphics[width=1.2\textwidth]{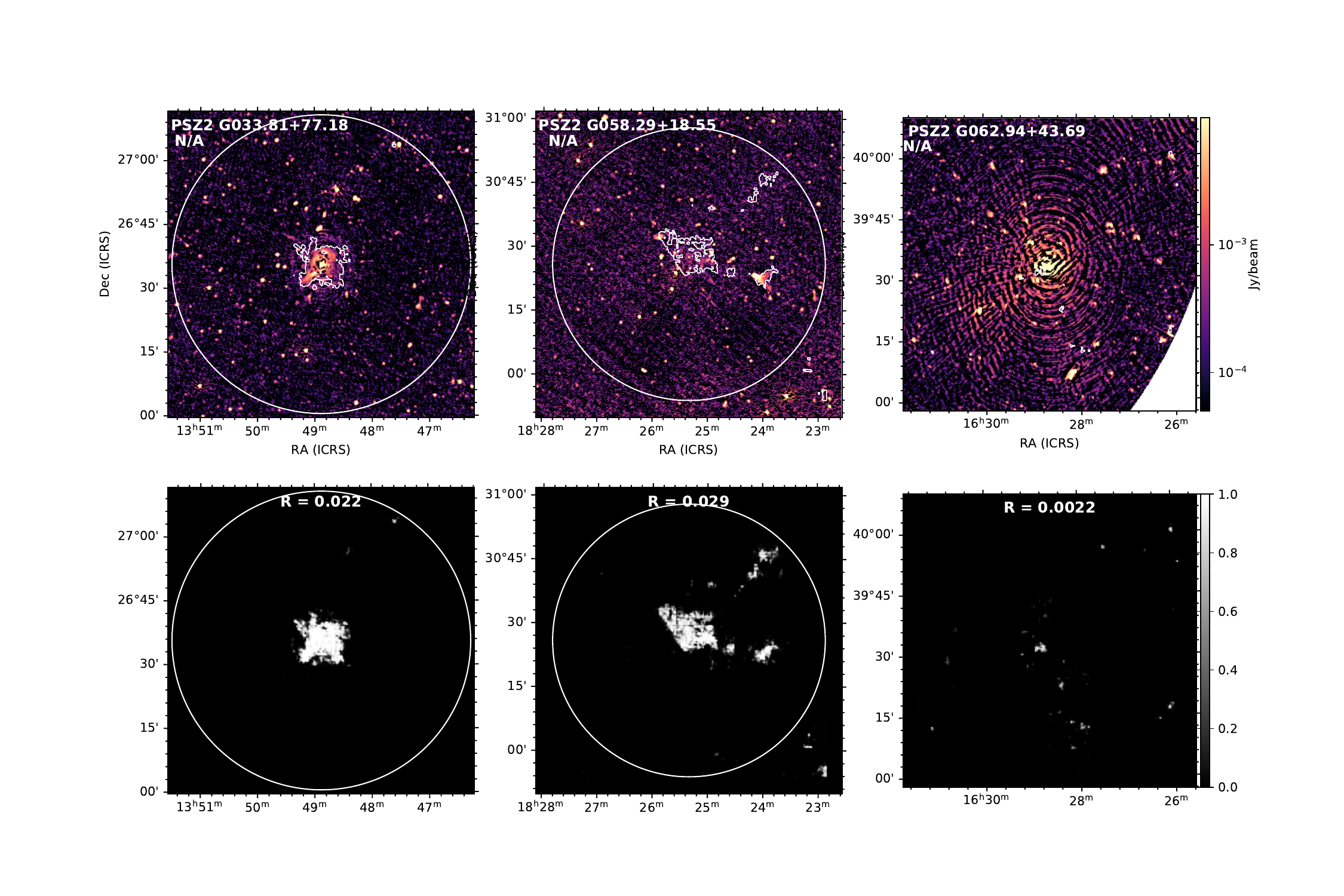}
\caption{Examples of LoTSS-DR2/PSZ2 galaxy clusters with low-quality images processed by Radio U-Net (see Sec.~\ref{sec:discussion}). Low-resolution ($20\arcsec$ beam) images of the LoTSS survey are shown in the first row while their segmented maps are shown in the bottom row.  The 0.5 probability contour is also reported in the images of the first row. The white circle shows the reference 2.2$R_{500}$ radius where $\mathcal{R}$ is computed. In the case of PSZ2 G062.94+43.69, the value of 2.2$R_{500}$ exceeds the image size and $\mathcal{R}$ is computed in the visible area. The left and central panels show two examples where Radio U-Net detected the presence of a radio halo (confirmed by targeted observations) while the galaxy cluster in the right panel is correctly classified as a non-detection.}
\label{fig:na}
\end{center}
\end{figure*}

In the previous Section, we applied Radio U-Net on a labeled sample of galaxy clusters and we discussed the detection statistics. Here, we focus on the investigation of the properties of wrong predictions, both false positives and false negatives, to understand in what circumstances the network fails. We then discuss the results. In the next Sections, we will refer to the results obtained on the initial test set, where we directly applied the network trained on synthetic observations.

\subsection{False negative: undetected sources}

Among the galaxy clusters hosting diffuse radio emission (DE and DEU), 23 over 132 are undetected. For the majority of them, R$_{500}$ has a small angular extent, both due to the small mass and/or high redshift. An example is represented by the right panel in Fig.~\ref{fig:bad_example_orig}.

Most false negative galaxy clusters show a region at the cluster's center with high detection probability, but the $\mathcal{R}$-value is too low to classify them as detected. This suggests that the performance of the network can be increased by improving our classification method, for instance, by adding a classification layer to the network.

It is interesting to note that not even one radio relic hosting system was missed, the sources in the false negative clusters being cataloged as RH (6), cRH (9), and U (8). This suggests that, by including radio halos in our simulations, in particular in low mass and high redshift systems, could potentially allow us to improve the model accuracy.

\subsection{False positive detections}

Considering the initial test set, 43 galaxy clusters classified as NDE in \citet{Botteon22}, host diffuse radio emission according to their $\mathcal{R}$ value. Clusters with false positive and true negative detections have similar distributions of redshift and angular extent of R$_{500}$. 

Most of the false detections are connected with the presence of radio galaxies with diffuse emission within the cluster, or by the convolution of radio emission from different radio galaxies (see left and central panel of Fig.~\ref{fig:bad_example_orig}). This is probably a consequence of the fact that the radio emission arising from galaxies, both extended and compact, is not present in the synthetic observations used for the training. However, the fact that the network identifies different kinds of diffuse radio emission is also encouraging because it means that it can generalize to sources never seen before.

Among the false positive detections, we also noticed some galaxy clusters for which the interesting region found by Radio U-Net is not restricted to extended radio galaxies. Two examples are shown in the left and central panel of Fig.~\ref{fig:fp}. In these cases, the regions spotted by the network are extended, covering a large fraction of the cluster area, and their morphology resembles that of radio halos. We speculate that a deeper observation with tailored data reduction and imaging procedures could lead to detect a diffuse radio emission in these systems. In particular, according to \citet{Cuciti22}, PSZ2 G073.97-27.82 is one of the galaxy clusters in the LoTSS-DR2/PSZ2 sample having the highest chance to host a mega-halo, and indeed the emission that we detect surpasses $R_{500}$. Another interesting case is PSZ2 G080.16+57.65 (see right panel of Fig.~\ref{fig:fp}), where Radio U-Net spotted both a relic and the central radio halo which was detected at lower resolution in \citet{vanWeeren21} but not confirmed in \citet{Botteon22}. This confirms that Radio U-Net can recover diffuse radio emission below the sensitivity limits of the observations.

\subsection{Low-quality images}

\citet{Botteon22} excluded from their analysis 31 galaxy clusters (marked as N/A) due to strong calibration artifacts or very high rms noise levels. Since the synthetic images used for the training include high noise levels and artifacts, Radio U-Net is expected to be effective for these low-quality images. 

Fig.~\ref{fig:na} shows three examples in which Radio U-Net achieves a proper segmentation of these low-quality images. The two clusters in the left and central panels are known to host radio halos thanks to previous targeted observations \citep{Giacintucci14,Botteon19} and are indeed detected by the network. The network gives instead a low probability of detection to pixels that are bright only due to imaging artifacts, such as in the right image, where the segmented map does not contain any relevant detection. These images provide additional evidence of the capability of Radio U-Net to quickly identify and select interesting targets, even in low-quality datasets, thereby identifying the fields that require more customized and time-intensive data analysis.

\section{Conclusions}
\label{sec:conclusions}

Radio interferometric images, derived from advanced image processing techniques, pose significant challenges for any data analysis solution due to the presence of both random noise and artifacts with vastly different statistical and morphological properties. Diffuse radio sources, such as radio halos and radio relics, are rare and particularly challenging to detect because of their low surface brightness and smooth boundaries. Their detection often requires computationally expensive procedures such as source subtraction. The anticipated massive data volumes in the coming decade from observatories like the SKA will necessitate software tools capable of effectively leveraging HPC solutions and achieving full automation, as interactive human intervention and supervision will become impractical. Machine learning solutions appear to be ideally suited to meet these requirements.

We have explored the use of a fully convolutional neural network to segment radio images and detect diffuse radio sources in large-area surveys. We trained a modified version of the classic U-Net architecture, named Radio U-Net, on synthetic LOFAR HBA observations of diffuse radio emission based on cosmological simulations. The network creates segmented images of the same size as the input image, assigning to each pixel a probability of being part of a diffuse radio source. The segmentation can be used to claim the detection and, at the same time, it provides additional information about the sources, e.g. extent or position within the cluster. 

The goal of this work is to verify the performance of Radio U-Net on real LoTSS data, by performing a binary classification of a manually labeled sample of galaxy clusters \citep{Botteon22}. We evaluated the precision, recall, and accuracy of the detections by computing the ratio $\mathcal{R}$ of the sum probability to the total number of pixels within 2.2$R_{500}$, taken as a reference value.

Our main achievements are summarized as follows:
\begin{itemize}
    \item Radio U-Net can detect and perform the segmentation of diffuse radio sources in large sets of images directly retrieved from the archive, quickly and in a fully automated manner.
    \item For the initial test set of 246 galaxy clusters, Radio U-Net reaches a maximum accuracy of $73\%$ by setting the detection threshold at $\mathcal{R}=0.015$. The probability images closely resemble the morphology of detected diffuse radio emission.
    \item  The recall at maximum accuracy is $83\%$, indicating that $17\%$ of clusters hosting diffuse radio emission are missed by Radio U-Net. These are primarily radio-halo hosting clusters with small angular sizes, which are underrepresented in the synthetic training dataset. 
    \item The majority of false detections are due to the presence of radio galaxies with extended and complex morphologies, which are not present in synthetic observations. This result indicates that, even though we tested Radio U-Net's performance on diffuse radio emission in galaxy clusters, it can also be used to detect and segment different kinds of diffuse radio emission, both within and beyond galaxy clusters.
    \item Radio U-Net correctly segments low-quality images affected by strong residual artifacts, identifying those containing diffuse radio emission.
    \item We attempted fine-tuning the network by re-training the last layers on a sub-sample of the dataset. This led to higher recall but lower precision, resulting in similar accuracy compared to the original model. This is likely because the sub-sample is too small to provide additional information to the model.
\end{itemize}

These results demonstrate that starting from a vast dataset of synthetic observations, as similar as possible to interferometric images, it is possible to transfer the learned information to real data. This is particularly relevant for the detection and classification of diffuse radio emission, where large training datasets are not yet available. Although deep learning is not ready to fully replace human data inspection, it can already be used as a complementary tool to facilitate radio-astronomical data exploitation. For instance, when working on a large sample of galaxy clusters, Radio U-Net can be used to make an initial selection of interesting targets from low-quality images, on which subsequent more tailored and computationally expensive data processing can be performed. Furthermore, taking advantage of the segmentation maps produced, Radio U-Net can be used to blindly search for diffuse radio emission in an unexplored region of the sky by comparing the position and morphology of the source with respect to other multi-wavelength tracers of clusters, bridges, or cosmic web filaments.

Radio U-Net was developed for LOFAR HBA observations, following the LoTSS specifications. However, its usage can easily be extended to other interferometers, provided that a tool to create synthetic observations from the sky model is available.

Further developments include the creation of more comprehensive simulations with radio emissions from galaxies, both Active Galactic Nuclei and star-forming galaxies. Another advance would be to modify the network to include a classification layer based on probability maps. This could enable further classification of diffuse radio sources into radio halos and radio relics based on their morphology and position relative to the cluster center. Another possibility is to include multi-wavelength data, such as X-ray and optical images, to handle comprehensive information about each galaxy cluster. The deep learning field is rapidly evolving, and new strategies can be considered such as self-supervised representation learning \citep{Ericsson22,Han23}.

These advancements underscore the transformative potential of deep learning in astronomical research, paving the way for more efficient and automated analysis of vast and complex datasets in the era of next-generation observatories.

\section*{Data Availability}
\label{data}
The sky and clean images used for this work are publicly available in FITS format at https://owncloud.ia2.inaf.it/index.php/s/IbFPlCCcPUresrr.

\section*{Acknowledgements}
This paper is supported by the Fondazione ICSC, Spoke 3 Astrophysics and Cosmos Observations. National Recovery and Resilience Plan (Piano Nazionale di Ripresa e Resilienza, PNRR) Project ID CN\_00000013 "Italian Research Center for High-Performance Computing, Big Data and Quantum Computing" funded by MUR Missione 4 Componente 2 Investimento 1.4: Potenziamento strutture di ricerca e creazione di "campioni nazionali di R\&S (M4C2-19)" - Next Generation EU (NGEU), and it's also supported by (Programma Operativo Nazionale, PON), ``Tematiche di Ricerca Green e dell'Innovazione". We acknowledge the CINECA award under the ISCRA initiative, for the availability of high performance computing resources and support. The HPC tests and benchmarks this work is based on, have been produced on the Leonardo Supercomputer at CINECA (Bologna, Italy) in the framework of the ISCRA programme IscrC\_RICK (project: HP10CDUNG6). 
F.V. has been supported by Fondazione Cariplo and Fondazione CDP, through grant n. Rif: 2022-2088 CUP J33C22004310003 for "BREAKTHRU" project.
We also acknowledge the usage of online storage tools kindly provided by the INAF Astronomical Archive (IA2) initiative (http://www.ia2.inaf.it). 
We thank Martin Hardcastle for setting up the cutout
service for low-resolution LoTSS DR2 images.

\bibliographystyle{mnras}
\bibliography{franco,franco2,my_bib}

\end{document}